\documentclass[aps,prd,10pt,nofootinbib,showpacs,showkeys,superscriptaddress,preprintnumbers,altaffilletter,floatfix]{revtex4-2}

\pdfoutput=1
\usepackage{soul}
\usepackage{graphicx}
\usepackage{microtype}
\usepackage{amsmath}
\usepackage{amssymb}
\usepackage{subfigure}
\usepackage{hyperref}
\usepackage{url}
\usepackage{xcolor}
\usepackage{color}
\usepackage{mathrsfs}
\usepackage{calrsfs}
\usepackage{amsfonts}
\usepackage{latexsym}
\usepackage{ragged2e}
\usepackage{epsfig}
\usepackage{textcomp}
\usepackage{phaistos}
\usepackage[utf8]{inputenc}
\makeatletter
\renewcommand\@makefnmark{\hbox{\@textsuperscript{\normalfont\color{purple}\@thefnmark}}}
\renewcommand\@makefntext[1]{%
  \parindent 1em\noindent
            \hb@xt@1.8em{%
                \hss\@textsuperscript{\normalfont\@thefnmark}}#1}
\makeatother

\usepackage{verbatim}
\usepackage{caption}
\DeclareCaptionJustification{justified}{\leftskip=0pt \rightskip=0pt \parfillskip=0pt plus 1fil}

\definecolor{vividviolet}{rgb}{0.62, 0.0, 1.0}
\definecolor{amaranth}{rgb}{0.9, 0.17, 0.31}
\definecolor{palatinateblue}{rgb}{0.15, 0.23, 0.89}
\definecolor{brightpink}{rgb}{1.0, 0.0, 0.5}
\definecolor{cornflowerblue}{rgb}{0.39, 0.58, 0.93}
\definecolor{deepcarminepink}{rgb}{0.94, 0.19, 0.22}
\definecolor{radicalred}{rgb}{1.0, 0.21, 0.37}
\definecolor{darkgreen}{rgb}{0.06 0.64, 0.43}
\definecolor{darkpink}{RGB}{231,84,128}

\hypersetup{ linktoc=all,
    colorlinks, linkcolor={palatinateblue},
    citecolor={brightpink}, urlcolor={amaranth}
}

\graphicspath{{Images/}}

\def\sideremark#1{\ifvmode\leavevmode\fi\vadjust{\vbox to0pt{\vss
 \hbox to 0pt{\hskip\hsize\hskip1em
 \vbox{\hsize1.5cm\tiny\raggedright\pretolerance10000
 \noindent #1\hfill}\hss}\vbox to8pt{\vfil}\vss}}}

\usepackage{graphicx} 
\usepackage{comment}

\usepackage{cancel}

\newcommand{\be}{\begin{equation}}
\newcommand{\ee}{\end{equation}}
\newcommand{\bea}{\begin{eqnarray}}
\newcommand{\eea}{\end{eqnarray}}

\newcommand{\bseq}{\begin{subequations}}
\newcommand{\eseq}{\end{subequations}}

\usepackage[]{xcolor}
\usepackage{hyperref}

\usepackage{amsmath}

\begin{document}

\title{
Degenerate and connection-dependent cosmological sectors in 
\texorpdfstring{$f(Q,C)$}< gravity
}

\author{Ismael \surname{Ayuso}}
\email{ismael.ayuso@ehu.eus}
\affiliation{Department of Physics \& EHU Quantum Center, University of the Basque Country UPV/EHU, Bilbao 48080, Spain}

\begin{abstract}

We investigate cosmological solutions in $f(Q,C)$ gravity formulated within symmetric teleparallel geometry, where gravitation is described by the nonmetricity scalar $Q$ and the boundary term $C$, related to the Ricci scalar of General Relativity through $\mathring{R}=Q+C$ in the absence of curvature and torsion. The symmetry requirements of FLRW spacetime admit three distinct realizations of the affine connection, leading in principle to three different cosmological sectors within the same theory. We show that the connection field equations play a crucial role in determining the cosmological dynamics. In their simplest realization, these equations impose a constraint that renders the theory dynamically equivalent to $f(\mathring{R})$ gravity, causing the cosmological background equations associated with the three connection realizations to coincide. This defines a degenerate cosmological sector of $f(Q,C)$ gravity, in which three a priori distinct geometric constructions converge to the same cosmological dynamics. We then consider the complementary class of genuinely nonequivalent $f(Q,C)$ models, for which the choice of connection becomes physically relevant. In this regime, the connection sector introduces additional dynamical degrees of freedom and gives rise to novel cosmological phenomenology absent in $f(\mathring{R})$ gravity.

\end{abstract}

\maketitle

\section{Introduction}

General Relativity (GR) has been a remarkably successful theory of gravity, accurately describing all tested gravitational phenomena at Solar–System scales. Nonetheless, when extrapolated to quantum or cosmological regimes, GR faces several challenges. Among the most prominent cosmological issues are the cosmological constant problem and, more recently, tensions among different data samples—such as the Hubble tension \cite{Abdalla:2022yfr, CosmoVerseNetwork:2025alb}, as well as indications from DESI data that suggest a deviation from a constant dark-energy component \cite{DESI:2025zgx}. While part of the proposed solutions rely on new phenomenological dark-energy models, others explore modifications of GR as the fundamental gravitational theory \cite{DiValentino:2021izs}. Consequently, the framework of modified gravity is vast, but careful study is needed to uncover cosmological models that better accommodate observational data than $\Lambda$CDM \cite{CANTATA:2021asi, Capozziello:2011et}.

The foundation of GR is the Einstein–Hilbert action. This employs the curvature scalar  $\mathring{R}$ constructed from the Levi-Civita connection, which is symmetric (torsionless) and metric compatible $(\nabla_\alpha g_{\mu\nu}=0)$. Thus, it is natural to ask what happens when these constraints on the connection are relaxed. As a result, the connection $\Gamma^{\alpha}_{\;\;\mu\nu}$ can be treated as a new degree of freedom, analogous to the metric $g_{\mu\nu}$. Accordingly, the action must also be varied with respect to the connection, giving rise to the metric–affine formalism. It is often stated that this formalism, when applied to the Einstein-Hilbert action with the curvature scalar $R$ defined with a general connection, leads to the Levi–Civita connection. However, for this to hold, one must assume that matter is only coupled to the metric and not to the connection, as well as impose either torsionlessness or metricity, as demonstrated in Ref. \cite{Clifton:2011jh}. Consequently, loosening the constraints on the connection results in new phenomenological features already in the simplest scenario.

It is possible to group these departures from the Levi-Civita connection into scalar quantities. The torsion scalar $T$ is constructed from the antisymmetric part of the connection, while the nonmetricity scalar $Q$ measures the extent to which the covariant derivative fails to preserve the metric tensor. However, there are formulations that are dynamically equivalent to GR using purely the torsion scalar,  or the nonmetricity scalar \cite{Maluf:2013gaa, BeltranJimenez:2017tkd}.  In such cases, GR is recast, but endowing the connection with alternative properties. This shift in perspective forms the basis of the so-called “geometrical trinity of gravity” \cite{BeltranJimenez:2019tjy}.

Still, these equivalences are realized only when flatness and metricity are imposed in the torsion-based formulation (teleparallel metric gravity), and when flatness and vanishing torsion are imposed in the nonmetricity-based formulation (symmetric teleparallel gravity). In these cases, the curvature scalar of GR can be rewritten as $\mathring{R}=T+B$ or $\mathring{R}=Q+C$, respectively, where $B$ and $C$ are boundary terms that do not contribute to the field equations.

These relations provide a natural starting point for constructing modified theories of gravity based on the properties of the connection. Indeed, the equivalence between theories built from $\mathring{R}$, $T$, and $Q$ holds only at the linear level and is generally lost when the action is promoted to an arbitrary function of these scalars. The reason is that the three scalars differ only by total divergence terms, which are dynamically irrelevant in the linear case but contribute nontrivially in nonlinear extensions. This opens the door to new phenomenology and cosmological scenarios driven by the dynamics of the connection. Consequently, theories such as $f(T)$ and $f(Q)$, as well as the more recent $f(T,B)$ and $f(Q,C)$ models, arise naturally.

In this work, we focus on Symmetric Teleparallel Gravity (STG) in cosmology. Consequently, the affine connection must be curvature-free and symmetric. It must also inherit the symmetries of FLRW spacetime, namely homogeneity and isotropy. These requirements lead to three fully independent compatible connections, each giving rise to distinct field equations. Two of these connections introduce a free function $\gamma(t)$ into the field equations, thereby enlarging the space of admissible cosmological solutions compared to the remaining one, which we denote as the trivial connection, since in the coincident gauge all its components vanish. 

The first investigations of $f(Q)$ models focused on the trivial connection, as it is the simplest of the three cases. This allows for the construction of cosmological models in which the acceleration of the Universe or dark energy is explained through the functional form of $Q$ \cite{BeltranJimenez:2019tme, Lazkoz:2019sjl, Boiza:2025xpn}, enabling subsequent comparison with observations \cite{Ayuso:2020dcu, Sahlu:2024pxk,Ayuso:2021vtj}. Similarly, $f(Q,C)$ models have recently begun to be explored in the context of the trivial connection. Relevant developments include \cite{Murtaza:2025wtw}, which presents Bianchi-I cosmology in $f(Q,C)$ gravity together with a dynamical-system analysis; \cite{Samaddar:2025xhi, Artola:2025fup, Sharif:2025esv}, which propose FLRW cosmological models and demonstrate their consistency with observational data; \cite{Sadatian:2024lub}, which investigates inflationary dynamics in this framework; and \cite{Junior:2024xmm}, which studies black-hole and regular black-hole solutions.

Models of $f(Q)$ gravity constructed with the non-trivial connections associated with $\gamma(t)$ have been investigated in \cite{De:2025swf, Narawade:2024pxb, Yang:2024tkw, Paliathanasis:2025hjw} together with analysis about its dynamical system \cite{Paliathanasis:2023nkb, Shabani:2023nvm, Dutta:2025fqw}. However, the full potential arising from the use of the three possible connections in $f(Q,C)$ models remains largely unexplored. Some initial steps in this direction include Ref.~\cite{Paliathanasis:2024yea}, where the free parameters necessary for the existence of an attractor accurately describing cosmic acceleration are analyzed; Ref.~\cite{Shabani:2024rky}, where de~Sitter solutions are presented for specific choices of the function, and \cite{Shabani:2025qxn} for a dynamical system analysis for simple power-law forms of $f(Q,C)$.

The aim of this work is to extend these recent lines of research to the three possible connections focusing on the affine connection field equation that guarantees the conservation of the energy-momentum tensor. This equation introduces a constraint that reduces the independent solution space of the three connections, causing them to converge to a single cosmological solution in the simplest scenario. This defines a degenerate cosmological sector of $f(Q,C)$ gravity, in which the three admissible connection branches lead to identical cosmological dynamics.

The paper is organized as follows. In Sec. \ref{STG}, we introduce Symmetric Teleparallel Geometry and, in particular, the $f(Q,C)$ theory. In Sec. \ref{Sec:Cosmologyf(QC)}, we develop the cosmological framework for the three different connections allowed by the cosmological symmetries and derive the corresponding field equations. We show that the connection field equation depends on an integration constant,when this constant is set to zero, the three branches collapse into the same cosmological dynamics, defining a degenerate cosmological sector. This sector is analyzed in Sec.~\ref{sec:deg-sol}, where we also investigate analytical cosmological solutions, focusing on power-law and exponential models, and show that they admit maximally symmetric solutions. In Sec. \ref{Sec:numerical solutions}, we study numerical solutions for which the equivalence between the three connections is broken, and show that a phenomenology similar to that of $\Lambda$CDM can be reproduced. Finally, we present our conclusions in Sec. \ref{Sec:conclusions}.

\section{SYMMETRIC TELEPARALLEL GRAVITY}\label{STG}

In the framework of symmetric teleparallel theory we consider an affine connection that is torsion-free and curvature-free. In addition, the gravitational  interaction is carried out by the nonmetricity (instead of the curvature of the Levi-Civita connection). Nonmetricity provides a measure of the failure of the metric to be covariantly conserved. This can be mathematically described through the nonmetricity tensor: 
\bea
Q_{\lambda\mu\nu}:=\nabla_\lambda g_{\mu\nu}\; .
\eea

Consequently, if nonmetricity does not vanish, lengths and angles are not preserved under parallel transport. In addition, any affine connection can be decomposed into the Levi--Civita connection, which is torsionless and metric compatible, plus additional contributions arising from torsion and nonmetricity (see Ref.~\cite{Jarv:2018bgs}). In the torsionless case, which is the one relevant for the present work, this decomposition simplifies and can be written as \cite{Jarv:2018bgs, BeltranJimenez:2019tjy}:
\bea
\Gamma^{\alpha}_{\mu\nu}=\mathring{\Gamma}^{\alpha}_{\mu\nu}+L^{\alpha}_{\;\;\mu\nu}\; \label{dec-gamma},
\eea
where $\mathring{\Gamma}^{\alpha}_{\mu\nu}$ is the Levi-Civita connection, and
\bea
L^{\alpha}_{\;\;\mu\nu}&=&\frac{1}{2}Q^{\alpha}_{\;\;\mu\nu}-Q^{\;\;\;\alpha}_{(\mu\;\;\;\nu)}
\eea
is the disformation that incorporates the nonmetricity of the connection. For a general connection, $\Gamma$, the Riemann curvature tensor is defined as:
\bea
R^{\alpha}_{\;\;\beta\mu\nu}(\Gamma)=2\partial_{[\mu}\Gamma^{\alpha}_{\;\;\nu]\beta}+2\Gamma^{\alpha}_{\;\;[\mu|\lambda|}\Gamma^{\lambda}_{\;\;\nu]\beta}\; .\label{riemann}
\eea

Upon adopting the decomposition of the connection introduced in Eq.~\eqref{dec-gamma}, the Riemann tensor can be expressed as:
\bea
R^{\alpha}_{\;\;\beta\mu\nu}(\Gamma)=\mathring{R}^{\alpha}_{\;\;\beta\mu\nu}+2\mathring{\nabla}_{[\mu}L^{\alpha}_{\;\;\nu]\beta}+2L^{\alpha}_{\;\;[\mu\mid\lambda\mid}L^{\lambda}_{\;\;\nu]\beta}\; .
\label{transformationRL}
\eea
By performing the appropriate index contractions of the Riemann tensor, one obtains the corresponding expression for the curvature scalar associated with the decomposed connection:
\bea
R\left(\Gamma\right)=\mathring{R}+\mathring{\nabla}_\alpha\left(Q^\alpha-\tilde{Q}^\alpha\right)+\frac{1}{4}Q^{\alpha\beta\gamma}Q_{\alpha\beta\gamma}-\frac{1}{2}Q^{\gamma\alpha\beta}Q_{\alpha\gamma\beta}
-\frac{1}{4}Q^\alpha Q_\alpha+\frac{1}{2}\tilde{Q}^\alpha Q_\alpha\; ,
\label{0RQ}
\eea
where:
\bea
Q_\alpha:=Q_{\alpha\mu}^{\;\;\;\;\mu} \;\;\;\;\; \;\;\;\;\;\tilde{Q}_\alpha:=Q_{\mu\alpha}^{\;\;\;\;\mu}\; .
\eea

Consequently, we can define two quantities. The first one is the nonmetricity scalar, which measures the nonmetricity \cite{DAmbrosio:2021pnd}:
\bea
Q:=-\frac{1}{4}Q_{\alpha\mu\nu}Q^{\alpha\mu\nu}+\frac{1}{2}Q_{\alpha\mu\nu}Q^{\mu\alpha\nu}+\frac{1}{4}Q_\alpha Q^\alpha-\frac{1}{2}Q_\alpha \tilde{Q}^\alpha\; ,\nonumber
\eea
and the second one is: 
\bea
C:=-\mathring{\nabla}_\alpha\left(Q^\alpha-\tilde{Q}^\alpha\right)\; ,
\eea
which is a boundary term~\cite{Shabani:2025qxn}. Accordingly, for a flat spacetime with vanishing curvature scalar, Eq.~\eqref{0RQ} can be recast as:
\bea
\mathring{R}=Q+C\; .
\eea

This equivalence allows one to rewrite the Einstein--Hilbert action in terms of the nonmetricity scalar $Q$, up to a boundary term that does not contribute to the field equations. In this way, the action of Symmetric Teleparallel Equivalent General Relativity (STEGR) naturally arises. Such an equivalence is lost when the actions are generalized to arbitrary functions of $Q$, $C$, and $\mathring{R}$.

Motivated by these previous analysis, the $f(Q,C)$ theory emerges with the action:
\bea
S=\int{d^4x\sqrt{-g}\left[\frac{1}{2\kappa}f(Q,C)+\mathcal{L}_M\right]}\; ,
\eea
where $\kappa=8\pi G$, $f(Q,C)$ is a general function of $Q$ and $C$, and $\mathcal{L}_M$ is the matter Lagrangian. Applying variations with respect to the metric, we find the equations~\cite{Shabani:2025qxn}:
\bea
-\frac{f}{2}g_{\mu\nu}+2P^\lambda_{\;\;\mu\nu}\nabla_\lambda(f_Q-f_C)+\left(\mathring{G}_{\mu\nu}+\frac{Q}{2}g_{\mu\nu}\right)f_Q+\left(\frac{C}{2}g_{\mu\nu}-\mathring{\nabla}_\mu\mathring{\nabla}_\nu+g_{\mu\nu}\mathring{\nabla}^\alpha\mathring{\nabla}_\alpha\right)f_C=\kappa T_{\mu\nu}\; , \label{fried-mod}
\eea
where $T_{\mu\nu}$ denotes the energy–momentum tensor. On the other hand, performing variations with respect to the connection, and in the absence of hypermomentum, yields the following field equations:
\bea
\mathcal{C}_\lambda\equiv\nabla_\mu\nabla_\nu\left[\sqrt{-g}\left(f_Q-f_C\right)P^{\mu\nu}_{\;\;\;\;\;\;\lambda}\right]=0\; .\label{con-eq}
\eea

In the linear case, i.e.\ $f(Q,C)=Q+C$, the theory reduces to STEGR, yielding the standard Friedmann equations of GR from Eq.~\eqref{fried-mod} together with the automatic satisfaction of Eq.~\eqref{con-eq}.

We conclude this section by recalling that the conservation of the energy--momentum tensor is governed by \cite{De:2023xua}:
\bea
\kappa \mathring{\nabla}_\mu T^{\mu}_{\;\;\nu}=2\nabla_\mu\nabla_\nu\left[\sqrt{-g}\left(f_Q-f_C\right)P^{\mu\nu}_{\;\;\;\;\;\;\lambda}\right]\; .
\eea

Therefore, the conservation of the energy--momentum tensor follows directly from the connection field Eq.~\eqref{con-eq}. This property plays an important role in $f(Q,C)$ gravity, since once both the metric and connection field equations are satisfied, no additional conditions are required to guarantee energy--momentum conservation. Consequently, throughout this work we consistently impose the connection equations together with the modified Friedmann equations when constructing cosmological solutions.

\section{Cosmology in \texorpdfstring{$f(Q,C)$}{f(Q,C)} gravity}
\label{Sec:Cosmologyf(QC)}

Having derived the field equations of $f(Q,C)$ gravity within the framework of symmetric teleparallel geometry, we now specialize to cosmology. To this end, we consider the Friedmann--Lemaître--Robertson--Walker (FLRW) metric, which describes a homogeneous and isotropic universe and is invariant under spatial rotations and translations:
\bea
ds^2=-N^2(t)dt^2+a^2(t)\left(dr^2+r^2 d\theta^2+r^2\sin^2\theta \, d\phi^2\right)\; ,
\eea
where $N(t)$ and $a(t)$ denote the lapse function and the scale factor, respectively. These symmetries can be described through its Killing vectors \cite{Hohmann:2019nat}:
\begin{align}
\mathcal{R}_x&=\left(0,0,\sin\phi,\frac{\cos\phi}{\tan\theta} \right)\; ,& \mathcal{T}_x&=\left(0,\sin\theta\cos\phi,\frac{\cos\theta\cos\phi}{r},-\frac{\sin\phi}{r\sin\theta}\right)\; ,\nonumber\\
\mathcal{R}_y&=\left(0,0,-\cos\phi,\frac{\sin\phi}{\tan\theta}\right)\; ,&\mathcal{T}_y&=\left(0,\sin\theta\sin\phi,\frac{\cos\theta\cos\phi}{r},\frac{\cos\phi}{r\sin\theta}\right)\; ,\nonumber\\
\mathcal{R}_z&=(0,0,0,-1)\; ,& \mathcal{T}_z&=\left(0,\cos\theta,-\frac{\sin\theta}{r},0\right)\; ,
\end{align}
where $\mathcal{R}_i$ generate rotations and $\mathcal{T}_i$ generate translations. Thus, the Lie derivatives of the metric with respect to these Killing vector fields vanish:
\bea
\mathcal{L}_{\xi_i} g_{\mu\nu}=0\; ,\label{liemetric}
\eea
where $\mathcal{L}_{\xi_i}$ is the Lie derivative with respect to the Killing vector $\xi_i$. However, to fully define the spacetime, in addition to the metric, we must introduce the connection. So far, the only restriction imposed on the connection has been that it is torsionless and curvature-free in order to remain within the STG framework. Nevertheless, if we want the symmetries of the FLRW metric to permeate the entire spacetime, we must require the connection to satisfy the same symmetries. Hence, Eq. \eqref{liemetric}, which is automatically satisfied for the components of the metric, must be imposed on the components of the connection:
\bea
\mathcal{L}_{\xi_i} \Gamma^\lambda_{\;\;\mu\nu}=0\; .
\eea

Imposing the cosmological symmetries and the symmetric teleparallel constraints yields a system of equations for the 64 connection components. The resulting system admits three inequivalent solution branches, which define three distinct cosmological connections \cite{Shabani:2025qxn,DAmbrosio:2021pnd}. We distinguish among these as $\Gamma^{\text{(I)}}$, $\Gamma^{\text{(II)}}$, and $\Gamma^{\text{(III)}}$. 

The components for the connection $\Gamma^{\text{(I)}}$ reads:
\begin{align}
    &\Gamma ^t {}_{\mu\nu} = \left(\begin{matrix}
        \gamma & 0 & 0 & 0 \\
        0 & 0 & 0 & 0 \\
        0 & 0 & 0 & 0 \\
        0 & 0 & 0 & 0
    \end{matrix} \right)\,, &\Gamma ^r{}_{\mu\nu} = \left( \begin{matrix}
        0 & 0 & 0 & 0 \\
        0 & 0 & 0 & 0 \\
        0 & 0 & - r & 0 \\
        0 & 0 & 0 & - r \sin ^2 \theta
    \end{matrix} \right)\,, \nonumber \\
    &\Gamma ^\theta {}_{\mu\nu} = \left(\begin{matrix}
        0 & 0 & 0 & 0 \\
        0 & 0 & \frac{1}{r} & 0 \\
        0 & \frac{1}{r} & 0 & 0 \\
        0 & 0 & 0 & -\cos \theta \sin \theta
    \end{matrix} \right)\,, &\Gamma ^\phi{}_{\mu\nu} = \left( \begin{matrix}
        0 & 0 & 0 & 0 \\
        0 & 0 & 0 & \frac{1}{r}\\
        0 & 0 & 0 & \cot \theta \\
        0 & \frac{1}{r} & \cot \theta & 0
    \end{matrix} \right)\, ,
\end{align}
where, throughout this paper, $\gamma$ denotes an arbitrary function of time. This connection is commonly known as the \emph{trivial connection}, since in the coincident gauge all its components vanish and the covariant derivative reduces to a partial derivative. As we show below, the function $\gamma$ enters only through the component $\Gamma^{t}_{\;\;tt}$ and drops out of the field equations. Consequently, it represents a pure gauge freedom rather than an independent dynamical degree of freedom.

The components for the connection $\Gamma^{\text{(II)}}$ reads:
\begin{align}
    &\Gamma ^t {}_{\mu\nu} = \left(\begin{matrix}
        \gamma + \frac{\dot{\gamma}}{\gamma} & 0 & 0 & 0 \\
        0 & 0 & 0 & 0 \\
        0 & 0 & 0 & 0 \\
        0 & 0 & 0 & 0
    \end{matrix} \right)\,, &\Gamma ^r{}_{\mu\nu} = \left( \begin{matrix}
        0 & \gamma & 0 & 0 \\
        \gamma & 0 & 0 & 0 \\
        0 & 0 & - r & 0 \\
        0 & 0 & 0 & - r \sin ^2 \theta
    \end{matrix} \right)\,, \nonumber \\
    &\Gamma ^\theta {}_{\mu\nu} = \left(\begin{matrix}
        0 & 0 & \gamma & 0 \\
        0 & 0 & \frac{1}{r} & 0 \\
        \gamma & \frac{1}{r} & 0 & 0 \\
        0 & 0 & 0 & -\cos \theta \sin \theta
    \end{matrix} \right)\,, &\Gamma ^\phi{}_{\mu\nu} = \left( \begin{matrix}
        0 & 0 & 0 & \gamma \\
        0 & 0 & 0 & \frac{1}{r}\\
        0 & 0 & 0 & \cot \theta \\
        \gamma & \frac{1}{r} & \cot \theta & 0
    \end{matrix} \right)\,,
\end{align}
where the dot denotes derivatives with respect to $t$.

Finally, the components for the connection $\Gamma^{\text{(III)}}$ reads: 

\begin{align}
    &\Gamma ^t {}_{\mu\nu} = \left(\begin{matrix}
        - \frac{\dot{\gamma}}{\gamma} & 0 & 0 & 0 \\
        0 & \gamma & 0 & 0 \\
        0 & 0 & \gamma \; r^2 & 0 \\
        0 & 0 & 0 &\gamma\; r^2 \sin ^2 \theta
    \end{matrix} \right)\,, &\Gamma ^r{}_{\mu\nu} = \left( \begin{matrix}
        0 & 0 & 0 & 0 \\
        0 & 0 & 0 & 0 \\
        0 & 0 & - r & 0 \\
        0 & 0 & 0 & - r \sin ^2 \theta
    \end{matrix} \right)\,, \nonumber \\
    &\Gamma ^\theta {}_{\mu\nu} = \left(\begin{matrix}
        0 & 0 & 0 & 0 \\
        0 & 0 & \frac{1}{r} & 0 \\
        0 & \frac{1}{r} & 0 & 0 \\
        0 & 0 & 0 & -\cos \theta \sin \theta
    \end{matrix} \right)\,, &\Gamma ^\phi{}_{\mu\nu} = \left( \begin{matrix}
        0 & 0 & 0 & 0 \\
        0 & 0 & 0 & \frac{1}{r}\\
        0 & 0 & 0 & \cot \theta \\
        0 & \frac{1}{r} & \cot \theta & 0
    \end{matrix} \right)\,.\label{connec2component}
\end{align}

It is important to note that, owing to the form of the connection, $\gamma$ cannot be set to zero for connections $\Gamma^{\text{(II)}}$ and $\Gamma^{\text{(III)}}$. These results arise purely from the geometric requirements of flatness, vanishing torsion, and compatibility with FLRW symmetry. As such, they are model-independent and remain valid for both $f(Q)$ and $f(Q,C)$ theories.

\subsection{Field equations}

We now compute the scalars $Q$ and $C$ and evaluate the field equations, Eqs.~\eqref{fried-mod} and \eqref{con-eq}, for each of the three admissible connections. From now on, we will set the lapse function $N(t) = 1$.  In addition, we assume that the energy-momentum tensor $T_{\mu\nu}$ takes the form of a perfect fluid:
\bea
T^\mu_{\;\;\nu}=\left(\rho+p\right)U^\mu U_\nu+p g^\mu_{\;\;\nu}\; ,
\eea
where $\rho$ denotes the energy density, $p$ the pressure  and $U_\mu$ the four velocity.

\subsubsection{Connection I}

For the connection $\Gamma^{\text{(I)}}$ the nonmetricity scalar and the boundary term read:
\bea
&&Q=-6 H^2\; ,\\
&&C=6(3H^2+\dot{H})\; ,
\eea
where $H$ is the Hubble function defined as $H\equiv\dot{a}/a$. The no null field equations are:
\bea
&&\frac{1}{2}f+\left(3H^2-\frac{Q}{2}\right)f_Q-\frac{1}{2}C f_C+3H\dot{f}_C=\kappa\rho\; ,\label{I-00}\\
&&-\frac{1}{2}f+\left(-3H^2-2\dot{H}+\frac{Q}{2}\right)f_Q+\frac{1}{2}C f_C-2 H\dot{f}_Q-\ddot{f}_C=\kappa p\; .\label{I-11}
\eea

The remaining field equations, including those obtained from variations with respect to the connection ($\mathcal{C}_\lambda = 0$), are identically satisfied. As anticipated, the parameter $\gamma$ appearing in the component $\Gamma^t_{\;\; tt}$ drops out of the field equations and therefore does not represent an independent dynamical degree of freedom. Consequently, for given matter profiles $\rho$ and $p$, the dynamics of connection I is completely determined by the choice of the function $f(Q,C)$.

A distinctive feature of connection I is that the connection field equations impose no additional constraints. As a result, the dynamics is governed solely by the modified Friedmann equations, while the conservation of the energy--momentum tensor is automatically satisfied. This contrasts sharply with the remaining connection branches, for which the connection equations play a nontrivial dynamical role.

The comparison with the modified Friedmann equations obtained in $f(T,B)$ gravity within the framework of \textit{metric teleparallelism}~\cite{Bahamonde:2021gfp} shows that both theories lead to identical cosmological equations (recall that $C=B$ in a flat FLRW background). Consequently, for this particular choice of connection, $f(Q,C)$ gravity does not exhibit any cosmological dynamics beyond those already explored in $f(T,B)$ models~\cite{Bahamonde:2015zma,Paliathanasis:2022pgu,Caruana:2020szx,Farrugia:2018gyz,Moreira:2021xfe}. Therefore, any genuinely new phenomenology in $f(Q,C)$ gravity must originate from alternative choices of the connection. Nevertheless, this branch remains of interest, as it may provide a different geometrical framework for investigating phenomenological scenarios that have not yet been studied within the context of $f(T,B)$ gravity.

\subsubsection{Connection II}

For the connection $\Gamma^{\text{(II)}}$, the nonmetricity scalar and the boundary term read \cite{Shabani:2025qxn}:

\bea
&&Q=-6 H^2+9\gamma H+3\dot{\gamma}\; ,\\
&&C=6(3H^2+\dot{H})-9\gamma H-3\dot\gamma\; .
\eea

The equations of motion that are not trivially satisfied in this case are:
\bea
&&\frac{1}{2}f+\left(3H^2-\frac{Q}{2}\right)f_Q-\frac{1}{2}C f_C+\frac{3}{2}\gamma\left(\dot{f}_Q-\dot{f}_C\right)+3H\dot{f}_C=\kappa\rho\; ,\label{II-00}\\
&&-\frac{1}{2}f+\left(-3H^2-2\dot{H}+\frac{Q}{2}\right)f_Q+\frac{1}{2}C f_C+\frac{3}{2}\gamma\left(\dot{f}_Q-\dot{f}_C\right)-2H\dot{f}_Q-\ddot{f}_C=\kappa p\label{II-11}\; ,\\
&&\mathcal{C}_t\equiv\frac{3}{2}\gamma\left(3H\dot{f}_Q+\ddot{f}_Q-3H\dot{f}_C-\ddot{f}_C\right)=0\; .\label{eq-conII}
\eea

Unlike connection I, in addition to the two field equations arising from the metric variations, one must also consider Eq.~\eqref{eq-conII}, which follows from variations with respect to the connection and is not automatically satisfied. This equation corresponds to the temporal component of Eq.~\eqref{con-eq}. As in \cite{Ayuso:2025vkc}, it can be written as:
\bea
3H(\dot{f}_Q-\dot{f}_C)+\frac{d}{dt}(\dot{f}_Q-\dot{f}_C)=0
\eea
with the solution:
\bea
\dot{f}_Q-\dot{f}_C=\frac{\alpha_0}{a^{3}}\label{CII}\; ,
\eea
where $\alpha_0$ is a constant. Therefore, the simplest case without considering a specific temporal evolution for the scalar factor is to take $\alpha_0=0$. Then, Eq. \eqref{CII} reads as $f_Q-f_C=k$ for a constant $k$, and the solution of $f(Q,C)$ becomes:
\bea
f(Q,C)=g(Q+C)+\frac{k}{2}(Q-C)\; ,\label{solCII}
\eea
where $g(Q+C)$ is an arbitrary function of $Q+C$. Let us remark that $Q+C$ is a special combination in which the dependence on $\gamma$ vanishes. In fact, $\gamma$ no longer contributes to the field equations. This conclusion is easily inferred by introducing solution \eqref{solCII} into \eqref{II-00} and \eqref{II-11}, resulting in expressions that contain only $Q+C$ terms, even for nonzero $k$. Consequently, setting $\alpha_0=0$ effectively shifts the degree of freedom associated with the connection to a spurious function without physical relevance.

Another interesting case arises for $\gamma = 2H(t)$. In this scenario, the dependence on $\alpha_0$ in the field equations for $\rho$ and $p$ disappears. Indeed, for this choice of $\gamma$, $C$ vanishes and, although $f_C$ need not vanish, Eq.~\eqref{II-00} reduces to the field equation for $\rho$ in an $f(Q)$ model with the same connection. However, this is not the case for the pressure equation, which, although independent of $\alpha_0$, differs from its $f(Q)$ counterpart due to additional modifications. This suggests that this case can reproduce cosmological evolutions analogous to those of $f(Q)$ gravity, but with a modified effective equation-of-state parameter $\omega_{\rm eff}$. The field equations for this case, after imposing the constraint in Eq.~\eqref{CII}, become:
\bea
&&\frac{1}{2}f+\left(3H^2-\frac{Q}{2}\right)f_Q+3H\dot{f}_Q=\kappa\rho\; ,\\
&&-\frac{1}{2}f+\left(-3H^2-2\dot{H}+\frac{Q}{2}\right)f_Q-2H\dot{f}_Q-\ddot{f}_Q=\kappa p\; ,
\eea
where $f$ remains a function of $Q$ and $C$, that is, $f = f(Q,C)$ \footnote{
The equation for the pressure of a $f(Q)$ model and connection II with $\gamma=2H$ reads:
\bea
-\frac{1}{2}f+\left(-3H^2-2\dot{H}+\frac{Q}{2}\right)f_Q+H\dot{f}_Q=\kappa p\;
\eea
}. 
However, although it does not enter explicitly into the matter equations, the constraint in Eq.~\eqref{CII} must still be satisfied, or equivalently checked for the chosen $f(Q,C)$. Thus, the extension from $f(Q)$ to $f(Q,C)$ is subject to this additional condition.

\subsubsection{Connection III}

It is possible to simplify the equations obtained for this case through a redefinition in the components of the connection such as $\gamma_*=\gamma/a^2$, in a manner analogous to \cite{Ayuso:2025vkc}. Then, for the connection $\Gamma^{\text{(III)}}$ the nonmetricity scalar and the boundary term reads:
\bea
&&Q=-6H^2+\frac{3}{a^2}\left(\gamma H+\dot{\gamma}\right)=3\left(3 H \gamma_*+\dot{\gamma_*}-2 H^2\right)\; ,\\
&&C=6\left(3H^2+\dot{H}\right)-\frac{3}{a^2}\left(\gamma H+\dot{\gamma}\right)=6\left(3H^2+\dot{H}\right)-3\left(3H\gamma_*+\dot{\gamma_*}\right)\; .
\eea

Let us remark that, after the redefinition, the expressions for $Q$ in connections II and III become identical. The same holds for the expressions for $C$. The field equations for this connection are:
\bea
&&\frac{1}{2}f+\left(3H^2-\frac{Q}{2}\right)f_Q-\frac{1}{2}C f_C-\frac{3}{2}\gamma_*\left(\dot{f}_Q-\dot{f}_C\right)+3H\dot{f}_C=\kappa\rho\; ,\label{III-00}\\
&&-\frac{1}{2}f+\left(-3H^2-2\dot{H}+\frac{Q}{2}\right)f_Q+\frac{1}{2}C f_C+\frac{1}{2}\gamma_*\left(\dot{f}_Q-\dot{f}_C\right)-2H\dot{f}_Q-\ddot{f}_C=\kappa p \; ,\label{III-11}\\
&&\mathcal{C}_t\equiv\frac{3}{2}\gamma_*\left(\left(5H+2\frac{\dot{\gamma}_*}{\gamma_*}\right)(\dot{f}_Q-\dot{f}_C)+(\ddot{f}_Q-\ddot{f}_C)\right)=0\; .\label{eq-CIII}
\eea

We can define: $X=a^5(\dot{f}_Q-\dot{f}_C)$. Then, Eq. \eqref{eq-CIII} becomes:
\bea
\dot{X}=-2X\frac{\dot{\gamma}_*}{\gamma_*}\; .
\eea
Consequently:
\bea
\dot{f}_Q-\dot{f}_C=\frac{\alpha_0}{a^5 \gamma_*^2}\; .\label{eq-solCIII}
\eea

As in connection II, setting $\alpha_0=0$ yields the constraint that $f(Q,C)$ must take the form given in Eq.~\eqref{solCII}. However, the choice $\gamma_* = 2H(t)$, while still implying $C=0$, does not remove all contributions proportional to $f_C$ from the equation for $\rho$. Consequently, unlike in connection II, the resulting equations do not reduce to those of an $f(Q)$ model. Interestingly, choosing $\gamma_* = -2H(t)$ does eliminate the terms proportional to $\dot{f}_C$ in the equation for $\rho$. Nevertheless, in this case $C$ no longer vanishes, and therefore the third term in Eq.~\eqref{III-00} remains nonzero.

Finally, a comparison of the field equations for the three connections reveals that Eq.~\eqref{III-00} differs from Eq.~\eqref{II-00} only by the sign of the term proportional to $\dot{f}_Q-\dot{f}_C$, while Eq.~\eqref{III-11} differs from Eq.~\eqref{II-11} by a factor of three in the same contribution. Therefore, whenever
\begin{equation}
\dot{f}_Q-\dot{f}_C=0,
\end{equation}
the field equations for connections II and III become identical. Since this contribution is absent for connection I [Eqs.~\eqref{I-00} and \eqref{I-11}], the equations for $\rho$ and $p$ coincide for all three connections. In particular, this occurs when $\alpha_0=0$.

At first sight, one might expect the resulting cosmological solutions to differ because the scalars $Q$ and $C$ are defined differently for each connection. However, when $\alpha_0=0$, the connection equation constrains the function $f(Q,C)$ to take the form given in Eq.~\eqref{solCII}. In this case, $Q+C$ is identical for all three connections, while the dependence on $Q-C$ enters the field equations only through the constant parameter $k$. Consequently, any function $f(Q,C)$ satisfying Eq.~\eqref{solCII} leads to the same cosmological solutions for all three connection branches. This defines a degenerate cosmological sector of $f(Q,C)$ gravity, in which three geometrically distinct connection branches give rise to identical cosmological dynamics.

Although this result is consistent with the fact that the theory effectively reduces to an $f(\mathring{R})$-type model, it is important to emphasize that the form of $f(Q,C)$ is not chosen arbitrarily. Rather, it is enforced by the connection equation, which imposes a strong restriction on the allowed functional dependence of $f(Q,C)$. The equivalence between the three branches is therefore broken only when $\alpha_0\neq0$, a situation that generally requires a nontrivial tuning between the function $f(Q,C)$ and the connection variable $\gamma$ (or $\gamma_*$) in order to satisfy Eq.~\eqref{CII} [or Eq.~\eqref{eq-solCIII}].

In the next two sections, we derive cosmological solutions in both regimes. We first study the degenerate cosmological sector corresponding to $\alpha_0=0$, and then turn to numerical solutions beyond this sector, where the degeneracy is lifted and the three connection branches lead to genuinely different cosmological dynamics.

\section{Analytical Cosmological Solutions in the Degenerate Sector}\label{sec:deg-sol}

In this section, we focus on the class of $f(Q,C)$ theories satisfying $\dot{f}_Q=\dot{f}_C$, a condition that ensures the connection field equations are satisfied. The general solution is given by:
\bea
f(Q,C)=g(Q+C)+\frac{k}{2}(Q-C)\; . 
\eea

As shown in the previous section, this condition makes the cosmological field equations associated with the three admissible connections identical and removes any dependence on the free connection components. This defines a degenerate cosmological sector of $f(Q,C)$ gravity, in which the three geometrically distinct connection branches give rise to identical cosmological dynamics and solutions.

It is important to emphasize that the condition $\dot{f}_Q=\dot{f}_C$ is not imposed by hand, but follows from the connection field equation $\mathcal{C}_t=0$ in the branch $\alpha_0=0$ [see Eqs.~\eqref{CII} and \eqref{eq-solCIII}]. Therefore, the degeneracy between the three connection branches is lifted only when $\alpha_0\neq0$. In what follows, we restrict our analysis to the energy-density equation, since the pressure equation is automatically satisfied by the continuity equation,
\bea
\dot{\rho} = -3H(\rho + p)\; ,
\eea
as can be verified explicitly.

\subsection{Polynomial solutions}

We consider the case
\bea
f(Q,C)=\alpha (Q+C)^n+\frac{k}{2}(Q-C)\; ,
\eea
where $\alpha$ is a constant that we set to unity , with dimensions $\kappa^{-(n-1)}$. This choice represents the simplest polynomial extension driven by the combination $Q+C$, while retaining the linear contribution proportional to $Q-C$. As discussed previously, models of this form satisfy the condition $\dot{f}_Q=\dot{f}_C$, ensuring that the three connection sectors share the same cosmological dynamics. Consequently, substituting this function into the field equations for any of the three connections yields the same expression for $\rho$. The resulting equation contains nontrivial corrections involving powers of $2H^2+\dot H$ together with derivative contributions, reflecting the higher-order character of the model:
\bea
H^2 \left[6^n n \left(2 H^2+\dot{H}\right)^{n-1}+3 k\right]-6^n
(n-1) \left(2 H^2+\dot{H}\right)^n
+6^n (n-1) n H \left(2 H^2+\dot{H}\right)^{n-2}
\left(4 H \dot{H}+\ddot{H}\right)
=2\kappa\rho\; .\label{eq-pol0}
\eea

At first sight, Eq.~\eqref{eq-pol0} appears rather involved due to the coexistence of nonlinear powers and derivative terms. However, its structure becomes considerably more transparent once the geometrical combinations entering $Q$ and $C$ are identified. In particular,
\begin{equation}
Q+C=6(2H^2+\dot H),
\qquad
\dot Q+\dot C=6(4H\dot H+\ddot H).
\end{equation}
The latter relation reveals an explicit higher-order structure, since
\begin{equation}
4H\dot H+\ddot H=\frac{d}{dt}(2H^2+\dot H),
\end{equation}
suggesting that the quantity $X\equiv2H^2+\dot H$ acts as an effective dynamical variable governing the cosmological evolution. Rewriting the field equations in terms of $X$ provides a more compact representation of the dynamics. Using these relations, Eq.~\eqref{eq-pol0} can be rewritten (provided that $Q+C\neq 0$) as
\bea
  n H^2 -
   (n-1) \left[\left(2 H^2+\dot{H}\right)- n H \frac{d}{dt}\log\left(2 H^2+\dot{H}\right)\right]
   =\frac{2\kappa\rho-3k H^2}{6^n\left(2 H^2+\dot{H}\right)^{n-1}}\; .\label{eq-pol}
\eea

Eq.~\eqref{eq-pol} makes the role of the parameter $n$ transparent, as deviations from standard cosmological evolution are encoded both in the nonlinear dependence on $2H^2+\dot H$ and in the logarithmic derivative term. The special case $n=1$ removes all higher-order contributions proportional to $(n-1)$, yielding a significantly simpler Friedmann-like dynamics, similar to that of General Relativity, albeit with a modified gravitational coupling constant:
\bea
\frac{1}{2}\left(6+3k\right)H^2 = \kappa\rho \, ,
\eea
recovering the standard General Relativity result in the limit $k=0$.

Having analyzed the general cosmological structure of the model, we now turn to a particular class of exact solutions. Maximally symmetric backgrounds provide a simple setting to investigate whether the theory naturally accommodates de Sitter configurations.

\subsubsection{Maximally symmetric solutions}

We focus on maximally symmetric solutions by imposing a constant Hubble parameter, $H=H_0$. In addition, we consider vacuum configurations with $\rho=0$ and $p=0$. Then, Eqs.~\eqref{II-00} and \eqref{II-11} take the same form:
\bea
6H_0^2 k-12^n H_0^{2n}(n-2)=0,
\eea
with three possible solutions:
\begin{itemize}
    \item $H_0=0$. This solution corresponds to a static spacetime with vanishing Hubble expansion, $a=\mathrm{const}$, and represents the trivial vacuum branch of the theory.
    \item For $H_0\neq0$ and $k\neq0$, the theory admits a nontrivial de Sitter branch with a constant expansion rate determined by the model parameters. 
    \bea
    H_0^2=\left(\frac{k}{3^{n-1}2^{2n-1}(n-2)}\right)^{\frac{1}{n-1}} ,
    \eea
    In this case, the modified geometric sector effectively generates an accelerated vacuum solution without requiring the introduction of an explicit cosmological constant. The dependence of $H_0$ on $n$ further indicates that the vacuum structure is highly sensitive to the nonlinear corrections encoded in the polynomial form of the function $f(Q,C)$.
    \item For $k=0$, a nontrivial de Sitter branch exists only for $n=2$, in which case the Hubble parameter remains undetermined. This indicates the presence of a degenerate vacuum structure, where the theory does not dynamically fix the expansion rate.
\end{itemize}

The above solutions reveal a nontrivial vacuum structure. In particular, for $k=0$ the case $n=2$ becomes special, admitting a degenerate family of de Sitter solutions with arbitrary values of $H_0$. On the other hand, for $k\neq0$, de Sitter solutions exist for generic values of $n$, except for the special case $n=2$. Therefore, the model naturally accommodates a de Sitter phase without introducing an explicit cosmological constant.

\subsection{Exponential corrections}

Having explored polynomial extensions of the form $(Q+C)^n$, it is natural to investigate whether the previous results persist for more general functional forms. In particular, exponential corrections represent one of the simplest nonpolynomial generalizations and frequently arise in modified gravity scenarios as a way to introduce nonlinear effects while preserving a compact analytical structure. 

Such models have also been studied in the context of $f(\mathring{R})$ gravity, where the exponential contribution is typically introduced as a correction to the cosmological constant to provide a dynamical dark energy sector~\cite{Odintsov:2017qif,Jaime:2012nj}. In fact, recent generalizations of the exponential scenario, in which the argument of the exponential is raised to a power~\cite{DOnofrio:2025cuk}, have been suggested to mildly alleviate the Hubble tension when confronted with cosmological observations, including the DESI DR2 BAO measurements~\cite{DESI:2025zgx}.

In the present case, we treat these terms as independent contributions to analyze their behavior separately. Then, we consider the model
\bea
f(Q,C)=Q+C +\alpha e^{-\beta(Q+C)}-2\Lambda\; ,
\eea
where $\beta$ is a constant introduced to render the argument of the exponential dimensionless, $\Lambda$ denotes the usual cosmological constant, and $\alpha$ is a constant that controls the exponential corrections and carries the same dimensions as $\Lambda$.

As in the polynomial case, this choice satisfies the condition $\dot{f}_Q=\dot{f}_C$, ensuring that the cosmological dynamics remains identical for the three connection sectors while allowing for qualitatively different higher-order corrections. The modified Friedmann equation reads:
\begin{equation}
    3 H^2-\Lambda+\frac{\alpha}{2} e^{-6 \beta  \left(2 H^2+\dot{H}\right)} \left\{1+6   \beta  \left[\dot{H}+H \left(H+24 \beta  H \dot{H}+6 \beta  \ddot{H}\right)\right]\right\}=\kappa\rho\; .
\end{equation}

Consequently, in the limits $\alpha \to 0$ or $\beta \to +\infty$, the exponential contribution becomes negligible and the model reduces to General Relativity with a cosmological constant. The same behaviour is recovered when $Q+C$ becomes sufficiently large, which is expected to occur in the early Universe due to its relation with $\mathring{R}$.

Therefore, the model allows one to recover an early-time cosmological evolution close to that of the $\Lambda$CDM scenario, while introducing modifications at late times. This feature may provide a potential avenue to alleviate the Hubble tension.

\subsubsection{Maximally symmetric solutions}

As for the polynomial case, we consider maximally symmetric solutions by imposing a constant Hubble parameter, $H=H_0$, in vacuum with $\rho=0$ and $p=0$. Then, Eqs. \eqref{II-00} and \eqref{II-11} takes the same form:

\begin{equation}
    3H_0^2+\frac{\alpha}{2}e^{-12 H_0^2\beta}\left(1+6H_0^2\beta\right)-\Lambda=0\; .\label{eq:mss_exp}
\end{equation}

Unlike the polynomial case, the Minkowski branch is no longer automatically recovered.  Indeed, setting $H_0=0$ leads to the condition
\begin{equation}
\Lambda=\frac{\alpha}{2},
\end{equation}
showing that a static vacuum solution only exists for a specific choice of the model parameters. This indicates that the exponential sector effectively contributes to the vacuum energy, modifying the structure of maximally symmetric solutions. In contrast, recovering the Minkowski vacuum in the absence of a cosmological constant requires the exponential correction to vanish, namely $\alpha=0$.

Moreover, the field equation can be rewritten as
\begin{equation}
3H_0^2=\Lambda-\frac{\alpha}{2}e^{-12\beta H_0^2}(1+6\beta H_0^2),
\end{equation}
which can be interpreted as an effective cosmological constant depending on the Hubble scale itself. In this sense, the exponential sector behaves as a dynamical geometric contribution capable of shifting the de Sitter structure of the model.

Finally, Eq.~\eqref{eq:mss_exp} also shows that vacuum de Sitter solutions without a cosmological constant, i.e. for $\Lambda=0$, are possible provided that $\alpha(1+6H_0^2\beta)<0$. Therefore, the exponential correction acts as an effective source capable of supporting de Sitter solutions even when the cosmological constant vanishes.

\section{Numerical Cosmological Solutions Beyond the Degenerate Sector}\label{Sec:numerical solutions}

To explore scenarios beyond the $\mathring{R}$-equivalent branch, we consider the model:
\bea
f(Q,C)=Q+\alpha Q C=(1+\alpha C)Q\; ,
\eea
where $\alpha$ is a constant with dimensions of $H^{-2}$. The theory reduces to GR for $\alpha=0$ and provides a simple framework for identifying the effects introduced by the modified gravity sector. Unlike the models discussed in the previous section, whose dynamics is governed by the combination $Q+C$, the present choice contains an explicit interaction between the geometric scalars through the mixed term $QC$. As a result, the theory generally departs from the $\mathring{R}$-equivalent branch and can give rise to genuinely new cosmological behaviour.

From a geometrical perspective, the factor $(1+\alpha C)$ acts as a background-dependent modification of the gravitational sector, making the effective gravitational coupling sensitive to the cosmological evolution. In this respect, the model shares some similarities with nonminimally coupled and scalar--tensor theories, where gravity is governed by an effective coupling that evolves dynamically~\cite{Ayuso:2019bnw,Fujii:2003pa,Clifton:2011jh}. However, in contrast to scalar--tensor theories, the coupling factor introduced here neither possesses independent scalar dynamics nor is gravitation governed by the standard Ricci scalar of GR.

Furthermore, because the correction depends explicitly on $C$, the model introduces an additional geometric scale into the cosmological dynamics. This may lead to departures from the standard cosmological evolution, while naturally recovering GR-like behaviour in regimes where $|C|$ becomes sufficiently small.

We consider a Universe filled with radiation, dust, and a cosmological constant, with Equations of State (EoS) given by $p=\omega\rho$, where $\omega = 1/3$, $\omega = 0$, and $\omega = -1$, respectively. Since the conservation of the energy-momentum tensor is guaranteed by the connection equation Eq.~\eqref{con-eq}, the energy density $\rho(a)$ can be written in the usual form of the $\Lambda$CDM model:
\bea
\rho(a)=\frac{3 H_0^2}{\kappa}\left(\frac{\Omega_r}{a^4}+\frac{\Omega_m}{a^3}+\Omega_\Lambda\right)\; ,
\eea
where, $\Omega_r$, $\Omega_m$ and $\Omega_\Lambda$ stand for current values of the fractional densities of the three cosmological fluids comprising our matter-energy lot. However, $\Omega_\Lambda$ and $\Omega_r$ are not treated as free parameters. Indeed, $\Omega_\Lambda$ is determined by imposing the normalization condition at the present epoch, while $\Omega_r$ can be computed from the current value of the Hubble parameter \cite{Dodelson:2003ft}.

To facilitate a comparison among models, we also consider the energy density within the  Chevallier-Polarski-Linder (CPL) parametrization, currently favored by recent DESI collaboration observations \cite{DESI:2025zgx}. In this parametrization, the energy density evolves as \cite{Scherrer:2015tra}:
\bea
\rho_{\text{DE}}(a)=\rho^0_{\text{DE}}a^{-3(1+w_0+w_a)}e^{-3 w_a(1-a)}\; ,
\eea
where $\rho^0_{\text{DE}}$ is the present-day energy density.

From this point onward, we adopt the values reported by the DESI collaboration~\cite{DESI:2025zgx}. For the $\Lambda$CDM scenario, we use $h_0=0.6817$ and $\Omega_m=0.3027$, whereas for the CPL parametrization we take $h_0=0.6751$, $\Omega_m=0.3114$, $w_0=-0.838$, and $w_a=-0.62$.

\subsection{Connection I}

For connection I, upon introducing the definitions of $Q$ and $C$ and rewriting the field equations in terms of the scale factor $a$, we obtain\footnote{For the sake of clarity, we have not explicitly written the dependence on the scale factor $a$ of $H(a)$, $\rho(a)$, and $p(a)$.}:
\bea
&&3 \left(H^2+36 \alpha  H^4\right)=\kappa \rho\; , \label{eq:rho-con1}\\
&&-H \left(2 a  H'+3 H\right)-\alpha  H \left(144 a H^2  H '+108 H^3\right)=\kappa p\; , \label{eq:p-con1}
\eea
where prime denotes a derivative with respect to the scale factor $a$. This cosmological model is analogous to that obtained in five-dimensional Einstein--Gauss--Bonnet gravity \cite{Gomez:2022qsi}. Note that the case $\alpha > 0$ is typically associated with ultraviolet (UV) corrections arising in extra-dimensional theories, whereas $\alpha < 0$ is commonly related to corrections emerging in Loop Quantum Cosmology \cite{Ashtekar:2015iza}. 

Moreover, since for this connection the field equations $\mathcal{C}_\mu = 0$ are automatically satisfied, the conservation of the energy--momentum tensor is ensured and the continuity equation can be consistently employed without the need for additional constraints. This can be explicitly verified by noting that the pressure Eq. \eqref{eq:p-con1} is identically satisfied from Eq. \eqref{eq:rho-con1} with the continuity equation.

To compare the predictions of this model with those of the $\Lambda$CDM and CPL parametrizations, we consider different values of $\alpha$ while fixing the remaining free parameters of the $f(Q,C)$ theory to their corresponding $\Lambda$CDM values. In addition, we introduce the following dimensionless quantities for convenience:
\begin{equation}
    h\equiv \frac{H[\text{km}\; \text{s}^{-1} \text{Mpc}^{-1}]}{100[\text{km}\; \text{s}^{-1} \text{Mpc}^{-1}]}\; ,\quad \quad \quad \tilde\alpha= \alpha[\text{km}\; \text{s}^{-1} \text{Mpc}^{-1}]^{-2}\times 100^2 [\text{km}\; \text{s}^{-1} \text{Mpc}^{-1}]^2 \; .\label{eq:redefinition}
\end{equation}

\begin{figure}
    \centering
    \includegraphics[width=0.65\linewidth]{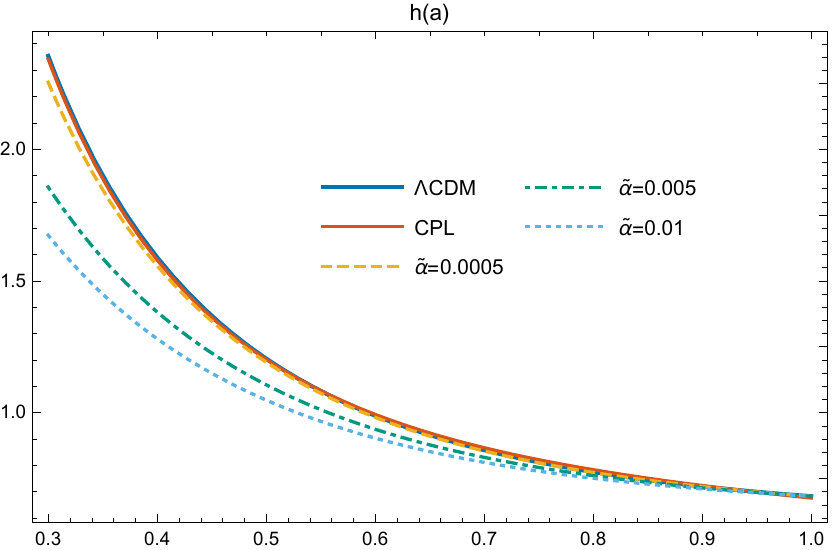}
    \caption{Cosmological evolution of the dimensionless Hubble function $h(a)$ for Connection I and several values of $\tilde{\alpha} $. We have also included the evolution for $\Lambda$CDM model and the CPL model. The matter density parameters adopted for the $f(Q,C)$ model are the same as those used in the $\Lambda$CDM scenario.} 
    \label{Connection1}
\end{figure}

The results are showed in Fig.~\ref{Connection1}, which displays the evolution of the dimensionless Hubble function $h(a)$ as a function of the scale factor for different positive values of $\tilde\alpha$, together with the $\Lambda$CDM and CPL models. As expected, all curves converge toward $h(a=1)=h_0$, where $h_0$ represents the present-day dimensionless value of the Hubble function, reflecting the normalization imposed at the current epoch.

At early times (small $a$), the deviations among the models are more noticeable. The modified gravity scenarios characterized by increasing values of $\tilde{\alpha}$ progressively depart from the $\Lambda$CDM prediction, leading to a different expansion rate in the past. In particular, larger values of $\tilde{\alpha}$ enhance the deviation from the standard cosmological behavior, whereas smaller values remain very close to $\Lambda$CDM. The figure further shows that, for the selected matter components, this case of the $f(Q,C)$ model yields a systematically smaller Hubble function than $\Lambda$CDM. This reduced expansion rate implies a slower cosmic evolution, typically resulting in a larger age of the Universe relative to the standard $\Lambda$CDM model.

The CPL parametrization exhibits an intermediate behavior, slightly departing from $\Lambda$CDM but remaining within the same qualitative trend. Overall, the figure illustrates that the parameter $\tilde{\alpha}$ controls the magnitude of the deviation from the standard cosmological expansion, allowing the model to interpolate between a $\Lambda$CDM-like evolution and more pronounced modified-gravity effects.

However, in the present analysis the matter sector, together with $h_0$ has been fixed to the values corresponding to the $\Lambda$CDM scenario to perform a qualitative comparison. Consequently, a fit of these parameters to cosmological data could improve the agreement between both models.

\subsection{Connection II}

For connection II, the field equations [Eqs. \eqref{II-00}-\eqref{eq-conII}]  in terms of the scale factor $a$, and using the expressions of $Q$ and $C$, becomes:

\bea
 &&2H^2+ 3 \alpha H^2 \left[\left(a \gamma'+3 \gamma\right)^2-2 H \left(a \left(\gamma'-a \gamma''\right)+15 \gamma\right)+24 H^2\right]+\frac{\alpha_0 \gamma }{a^3}=\frac{2}{3}\kappa\rho\; ,\label{eq-rho(a)II}\\ 
 &&\frac{3 \alpha_0 \gamma }{2 a^3}-2 a H H'-3 H^2+\alpha  \left\{-3 a^2 H \left(a \gamma'+3 \gamma\right) H'^2-3 H^3 \left[a \left(a^2 \gamma'''+4 a \gamma''-19 \gamma'+48  H'\right)-45 \gamma\right]\right.\nonumber\\
   &&\;\;\;\;\;\;\left.-\frac{3}{2} a^2 H^2 \left[6 a \gamma'' H'+\gamma' \left(3
   \gamma'+2 a H''+12 H'\right)\right]-9 a \gamma H^2 \left(3 \gamma'+a H''-6 H'\right)-\frac{81}{2} \gamma^2 H^2-108
   H^4\right\}=\kappa p\; ,\label{eq-p(a)II}\\
&&6 a \alpha  H \left\{ H' \left[a \left( H'-\gamma'\right)-3 \gamma\right]+H \left(-a \gamma''-4 \gamma'+a H''+9
    H'\right)\right\}=\frac{\alpha_0}{a^3}\label{eq-con(a)II}\; .
\eea

Let us recall that in these equations $H$, $\gamma$, $\rho$, and $p$ are functions of the scale factor $a$—whose explicit dependence has been omitted for clarity—whereas $\alpha$, $\alpha_0$ and $\kappa$ are constants. From Eq. \eqref{eq-con(a)II}, we observe that $\alpha = 0$ implies $\alpha_0 = 0$, thus recovering the GR limit and the equivalence among the three connections. As in Connection I, we rewrite the field equations in dimensionless form to facilitate the comparison with observational benchmarks and simplify the numerical analysis. Indeed, the previous system remains invariant under the substitutions given in Eq.~\eqref{eq:redefinition}, supplemented by the additional redefinitions:
\bea
\tilde\gamma=\frac{\gamma[\text{km}\; \text{s}^{-1} \text{Mpc}^{-1}]}{100[\text{km}\; \text{s}^{-1} \text{Mpc}^{-1}]}, \;\;\;\;  \tilde\alpha_0=\frac{\alpha_0[\text{km}\; \text{s}^{-1} \text{Mpc}^{-1}]}{100[\text{km}\; \text{s}^{-1} \text{Mpc}^{-1}]}\; \label{eq:redefinition2}.
\eea

Note that Eq.~\eqref{eq-rho(a)II} is lineal in $\gamma''$ (or equivalently in $\tilde\gamma''$). Consequently, we can solve its expression $\gamma''$ and substitute it into Eq.~\eqref{eq-con(a)II}. Simultaneously, we can redefine the functions as follows:
\bea
x_1(a)=h(a), \;\;\;\;\;\; x_2(a)= h'(a), \;\;\;\;\;\;x_3(a)=\tilde\gamma(a), \;\;\;\;\;\; x_4(a)=\tilde\gamma'(a)\; .
\eea

Analogously, Eq.~\eqref{eq-con(a)II} can be solved for $H''$. In this way, the original system formed by Eqs.~\eqref{eq-rho(a)II} and \eqref{eq-con(a)II} can be recast as a system of four first-order ordinary differential equations,
\bea
x'_1(a)=x_2(a), \;\;\;\;\;\; x'_2(a)=F(a,x_1,x_2,x_3,x_4), \;\;\;\;\;\; x'_3(a)=x_4(a), \;\;\;\;\;\;  x'_4(a)=G(a,x_1,x_2,x_3,x_4)\; ,\label{dyn-system}
\eea
where $F$ and $G$ are known functions. The resulting system can then be integrated numerically. Furthermore, the solution automatically satisfies Eq.~\eqref{eq-p(a)II} for the equation of state, as can be verified explicitly.

To obtain numerical solutions close to the $\Lambda$CDM background, we impose the initial conditions $h_0=h_0^{\Lambda \mathrm{CDM}}$ and $h'_0=(h'_0)^{\Lambda\mathrm{CDM}}$. Alternative choices, such as matching the CPL parametrization, are also possible and would naturally lead to solutions closer to CPL. Nevertheless, the parameters $(\tilde\alpha,\tilde\alpha_0,\tilde\gamma(a_0),\tilde\gamma'(a_0))$ remain free.

\begin{figure}
    \centering
    \includegraphics[width=0.45\linewidth]{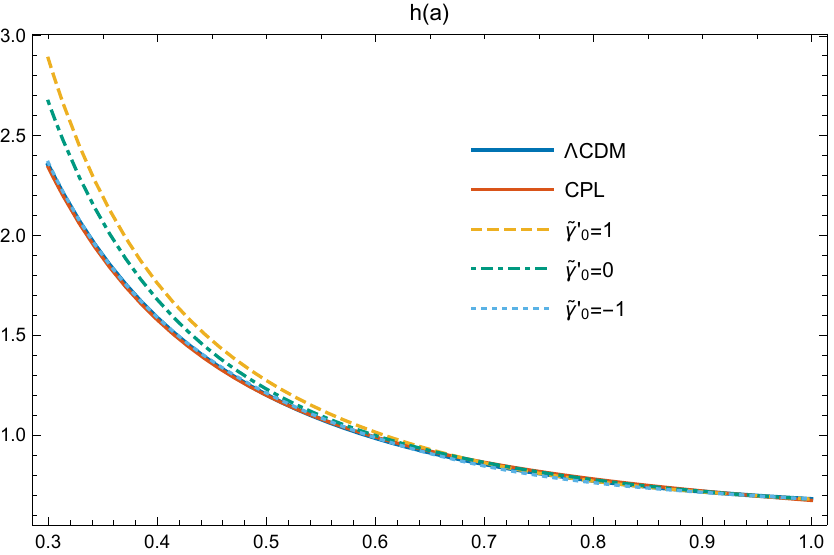}
    \includegraphics[width=0.455\linewidth]{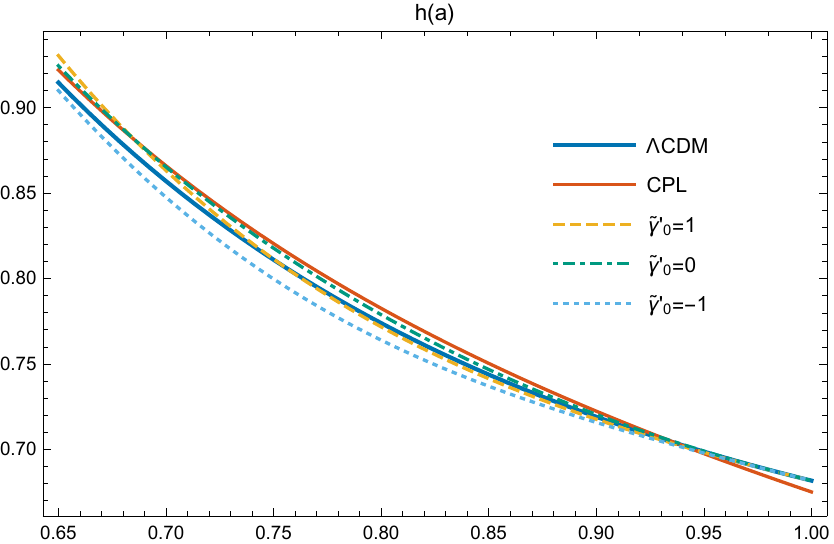}
    \centering
    \includegraphics[width=0.45\linewidth]{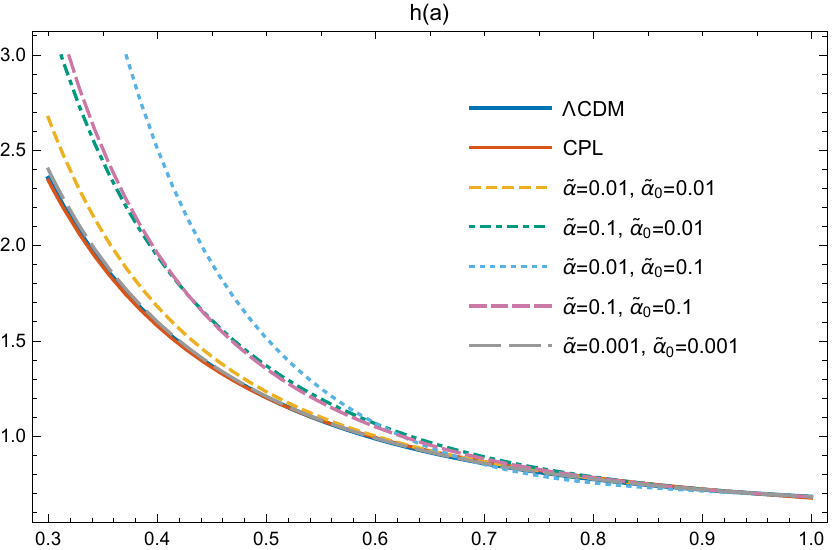}
    \includegraphics[width=0.455\linewidth]{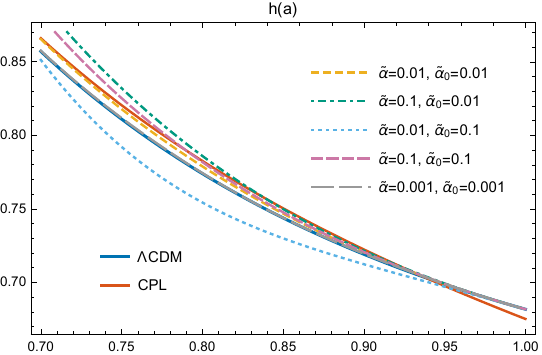}
    \caption{ Numerical solutions for the Hubble function and Connection II obtained by fixing the initial conditions at the present epoch, corresponding to $a_0=1$. The matter density parameters adopted for the $f(Q,C)$ model are the same as those used in the $\Lambda$CDM scenario. Top panels: the parameters are fixed to $\tilde\alpha = 0.01$, $\tilde\alpha_0 = 0.01$, and $\tilde\gamma_0\equiv\tilde\gamma(a_0) = 1$, while different values of $\tilde\gamma'_0\equiv\tilde{\gamma}'(a_0)$ are considered. Bottom panels: $\tilde\gamma_0 = 1$ and $\tilde\gamma'_0 =0$ are kept fixed, and different values of $\tilde\alpha$ and $\tilde\alpha_0$ are explored. }
    \label{fig:connectionII}   
\end{figure}

Fig.~\ref{fig:connectionII} shows several numerical solutions obtained by fixing the initial conditions at the present epoch, $a_0=1$. The upper panels illustrate the effect of varying $\tilde\gamma'_0\equiv\tilde\gamma'(a_0)$ while keeping the remaining parameters fixed at $\tilde\gamma_0\equiv\tilde\gamma(a_0)=1$, $\tilde\alpha=0.01$, and $\tilde\alpha_0=0.01$. In particular, the solution corresponding to $\tilde\gamma'_0=-1$ remains remarkably close to both the $\Lambda$CDM and CPL backgrounds, suggesting that an appropriate choice of parameters may yield an even better agreement with observations.

The lower panels display the impact of varying $(\tilde\alpha,\tilde\alpha_0)$ while fixing $\tilde\gamma_0=1$ and $\tilde\gamma'_0=0$. This choice is motivated by the requirement of obtaining cosmological evolutions close to $\Lambda$CDM. The resulting solutions for $h(a)$ remain phenomenologically viable and closely track the $\Lambda$CDM and CPL predictions. Nevertheless, they systematically exhibit a larger Hubble rate at earlier times. This trend can be reversed by modifying $\tilde\gamma'_0$, although at the expense of altering the late-time evolution. These results motivate a dedicated observational analysis in which all relevant parameters are allowed to vary, namely $\Omega_m$, $\tilde\alpha$, and $\tilde\alpha_0$, together with the initial conditions $h_0$, $h'_0$, $\tilde\gamma_0$, and $\tilde\gamma'_0$.

Finally, besides the choice of the initial conditions themselves, the epoch at which they are imposed also plays an important role. Throughout this section, the initial conditions have been fixed at the present epoch in order to keep the Hubble rate of the $f(Q,C)$ model as close as possible to the predictions of standard cosmological scenarios. However, one may instead specify the initial conditions at earlier stages of the cosmological evolution and study the subsequent dynamics of the model.

\begin{figure}
    \centering
    \includegraphics[width=0.45\linewidth]{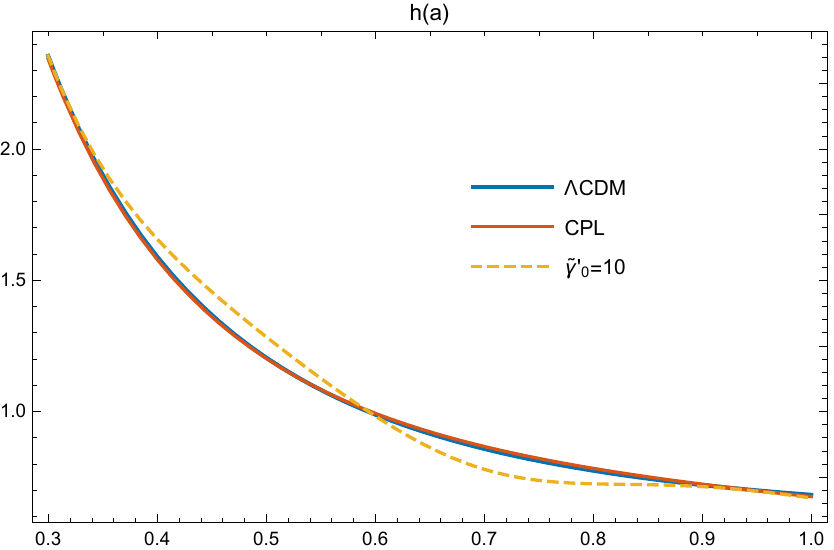}
   \includegraphics[width=0.455\linewidth]{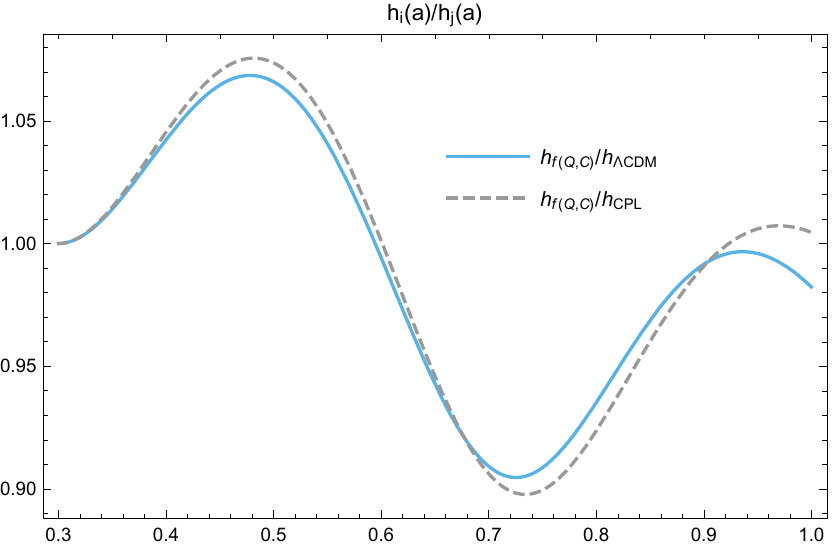}
    \caption{
    Numerical solutions for Connection II obtained by imposing the initial conditions at the earlier epoch $a_0=0.3$, with $\tilde{\gamma}_0=1$, $\tilde\alpha=0.01$,  $\tilde\alpha_0=0.01$ and $\tilde{\gamma}'_0=10$ in the $f(Q,C)$ model. For the right panel and the $f(Q,C)$ model, we adopt the corresponding values of $h_0$ from either the $\Lambda$CDM or CPL scenarios, depending on the reference model used for comparison.}
    \label{fig:hh}   
\end{figure}

This behaviour is illustrated in Fig.~\ref{fig:hh}, where both positive and negative deviations from the reference evolution of $h(a)$ can arise. These results further support the conclusion that a global fit of the free parameters may yield cosmological histories that closely mimic the $\Lambda$CDM or CPL backgrounds while retaining small deviations that could help alleviate existing cosmological tensions.

\subsection{Connection III}

For this connection, the field equations [Eqs.~\eqref{III-00}--\eqref{eq-CIII}], expressed in terms of the scale factor $a$ and using the definitions of $Q$ and $C$, take the form: 

\bea
 &&2H^2+ 3 \alpha H^2 \left[\left(a \gamma'_*+3 \gamma_*\right)^2-2 H \left(a \left(\gamma'_*-a \gamma''_*\right)+15 \gamma_*\right)+24 H^2\right]-\frac{\alpha_0  }{a^5 \gamma_*}=\frac{2}{3}\kappa\rho\label{eq-rho(a)III}\; ,\\
   &&\frac{ \alpha_0 }{2 a^5\gamma_*}-2 a H H'-3 H^2+\alpha  \left\{-3 a^2 H \left(a \gamma'_*+3 {\gamma_*}\right) H'^2-3 H^3 \left[a \left(a^2 \gamma'''_*+4 a \gamma''_*-19 \gamma'_*+48  H'\right)-45 {\gamma_*}\right]\right.\nonumber\\
   &&\;\;\;\;\left.-\frac{3}{2} a^2 H^2 \left[6 a \gamma''_* H'+\gamma'_* \left(3
   \gamma'_*+2 a H''+12  H'\right)\right]-9 a {\gamma_*} H^2 \left(3 \gamma'_*+a H''-6  H'\right)-\frac{81}{2} {\gamma_*}^2 H^2-108
   H^4\right\}=\kappa p\label{eq-p(a)III}\; ,\nonumber\\ \\
&&6 a \alpha  H \left\{ H' \left[a \left( H'-\gamma'_*\right)-3 {\gamma_*}\right]+H \left(-a \gamma''_*-4 \gamma'_*+a  H''+9
   H'\right)\right\}=\frac{\alpha_0}{a^5\gamma_*^2}\label{eq-con(a)III}\; .
\eea

As discussed above, the equations for $\rho$ and $p$ associated with this connection differ from those of Connection II only through the terms proportional to $\alpha_0$. At first sight, one might expect the two formulations to be equivalent at the level of the energy-density equation, since a sign change in $\alpha_0$ could be absorbed into a redefinition of the integration constant. However, this apparent equivalence is lost once the connection equation, Eq.~\eqref{eq-con(a)III}, is taken into account. Consequently, the two branches become fully equivalent only in the special case $\alpha_0=0$.

\begin{figure}
    \centering
    \includegraphics[width=0.45\linewidth]{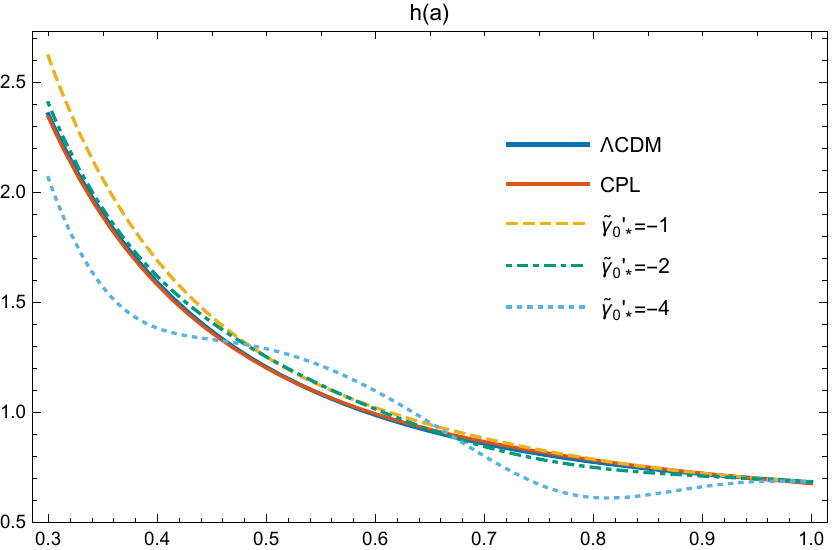}
    \includegraphics[width=0.455\linewidth]{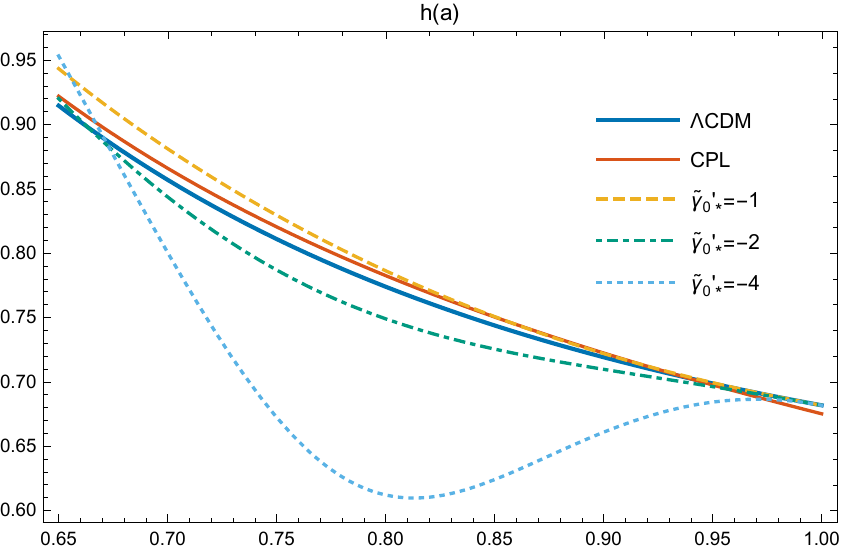}
    \centering
     \includegraphics[width=0.45\linewidth]{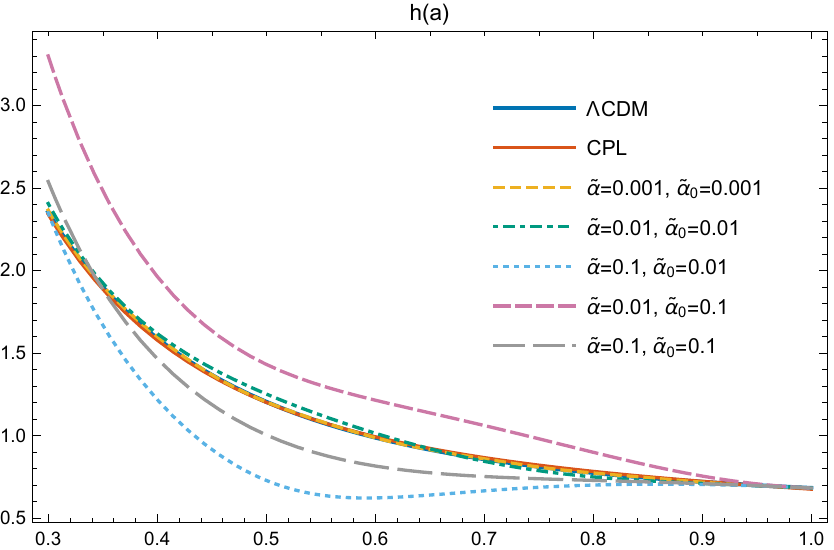}
    \includegraphics[width=0.455\linewidth]{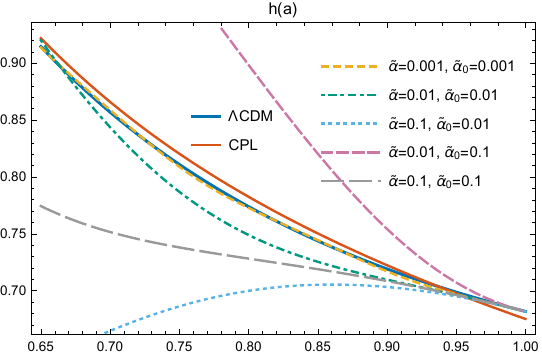}
    \caption{
    Numerical solutions for Connection III. Top panels: the parameters are fixed to $\tilde\alpha = 0.01$, $\tilde\alpha_0 = 0.01$, and $\tilde\gamma_{*0} = 1$, while different values of $\tilde{\gamma}'_{*0}$ are considered. Bottom panels: $\tilde\gamma_{*0}  = 1$ and $\tilde{\gamma}'_{*0}= -2$ are kept fixed, and different values of $\alpha_*$ and $\alpha_0^*$ are explored. Note that the green curves in all four panels correspond to the same solution and are included to facilitate the comparison of the remaining solutions across the different plots.
    }
    \label{fig:connectionIII}   
\end{figure}

Following the same procedure as for Connection II, we introduce the redefinitions given in Eqs.~\eqref{eq:redefinition} and \eqref{eq:redefinition2}, replacing $\gamma$ by $\gamma_*$. These transformations leave the system invariant and allow the equations to be written in dimensionless form in terms of the variables and parameters $(h,\tilde{\alpha},\tilde{\alpha}_0,\tilde{\gamma}_*)$. The resulting system can again be recast in the form of Eq.~\eqref{dyn-system}. To obtain solutions close to the $\Lambda$CDM background, we impose the initial conditions $h_0=h_0^{\Lambda\mathrm{CDM}}$ and $h'_0=(h'_0)^{\Lambda\mathrm{CDM}}$, leaving $(\tilde\alpha,\tilde\alpha_0,\tilde\gamma_*(a_0),\tilde\gamma'_*(a_0))$ as free parameters. The corresponding numerical solutions are presented in Fig.~\ref{fig:connectionIII}.

In the upper panels of Fig.~\ref{fig:connectionIII}, we fix $\tilde\alpha=0.01$, $\tilde\alpha_0=0.01$, and $\tilde\gamma_{*0}=1$ at the present epoch, $a_0=1$, while varying $\tilde\gamma'_{*0}$. Positive values of $\tilde\gamma'_{*0}$, as well as sufficiently large negative values, were found to produce divergences in $h(a)$. Accordingly, only values leading to regular cosmological evolutions are displayed.

The most distinctive feature of these solutions is the emergence of oscillations in $h(a)$ around the $\Lambda$CDM and CPL backgrounds. Their amplitude increases as $\tilde\gamma'_{*0}$ decreases, provided the divergent regime is avoided. To illustrate this behaviour, we include the case $\tilde\gamma'_{*0}=-4$, although the same qualitative trend is already visible for $\tilde\gamma'_{*0}=-2$, as shown in the upper-right panel.

These oscillations generate both positive and negative deviations from the standard cosmological evolution, depending on the epoch under consideration. As a result, the model is capable of producing departures from the $\Lambda$CDM background at different stages of the expansion history. Nevertheless, when the solutions are extrapolated to the recombination era ($z\approx1089$) and to neighbouring regions of the chosen parameter space, the Hubble rate typically becomes significantly larger than the corresponding $\Lambda$CDM and CPL predictions. This suggests that the oscillatory behaviour alone is unlikely to provide a substantial alleviation of the tension between early- and late-time cosmological observations.

In the lower panels, we fix $\tilde\gamma_{*0}=1$ and $\tilde\gamma'_{*0}=-2$ at the present epoch, $a_0=1$, while varying the pair $(\tilde\alpha,\tilde\alpha_0)$. As expected, small values of both parameters yield cosmological evolutions that closely track the $\Lambda$CDM background. This behaviour is consistent with the GR limit, since setting either parameter to zero forces the other to vanish as well. An interesting consequence of this correlation is that the departure from GR does not necessarily grow when both parameters are increased simultaneously. Instead, the largest deviations arise when one parameter is enhanced while the other is kept fixed. This is illustrated by the case $(\tilde\alpha,\tilde\alpha_0)=(0.1,,0.1)$, whose evolution remains considerably closer to the reference cosmological models than those corresponding to $(0.01,,0.1)$ or $(0.1,,0.01)$.

\begin{figure}
    \centering
    \includegraphics[width=0.65\linewidth]{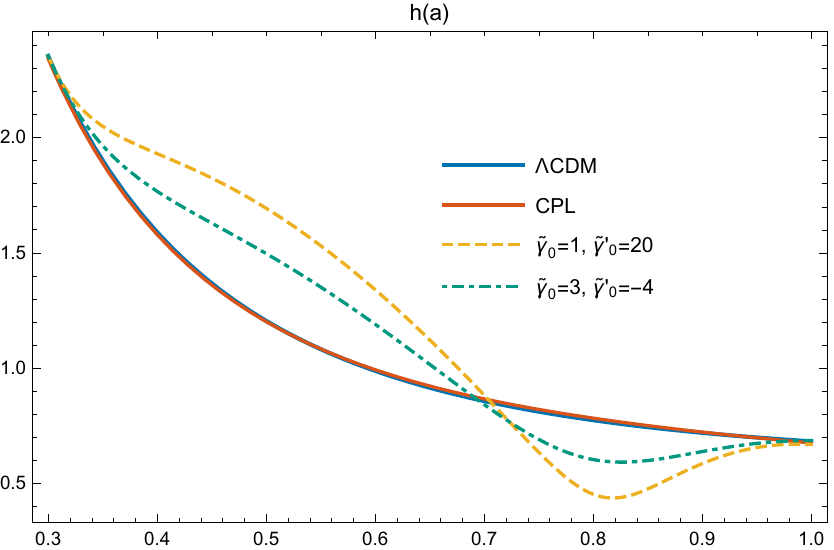}
    \caption{Numerical solutions for Connection III obtained by imposing the initial conditions at the earlier epoch $a_0=0.3$, with different values of $\tilde{\gamma}_{0*}$ and $\tilde{\gamma}'_{0*}$ in the $f(Q,C)$ model with $\tilde\alpha=0.01$, and $\tilde\alpha_0=0.01$. }
    \label{fig:con3_ind}   
\end{figure}

For completeness, and following the analysis of Connection II, Fig.~\ref{fig:con3_ind} shows the solutions obtained when the initial conditions are imposed at an earlier epoch, $a_0=0.3$, instead of the present time. In this case, the deviations in the Hubble rate are noticeably larger than those found for Connection II when $\tilde\gamma_{*0}=1$ is held fixed (yellow curve). These deviations can be partially reduced by varying $\tilde\gamma_{*0}$ as well (green curve). Nevertheless, compared with Connection II, the sign change in the term proportional to $\dot f_Q-\dot f_C$ appears to systematically amplify the departure from the $\Lambda$CDM and CPL backgrounds.

\section{Conclusions}\label{Sec:conclusions}

In this work, we have investigated symmetric teleparallel $f(Q,C)$ gravity in the context of FLRW cosmology. The symmetry requirements imposed on the connection lead to three distinct possibilities for its components. Furthermore, treating the connection as an independent dynamical degree of freedom gives rise to additional field equations whose consistency guarantees the conservation of the energy--momentum tensor. As a consequence, the cosmological dynamics is described by three different systems, each described by a set of modified Friedmann equations together with the corresponding equation of motion for the connection.

The first possible connection, referred to as Connection I, possesses the remarkable property that the connection field equations are automatically satisfied. As a result, the conservation of the energy--momentum tensor is guaranteed without imposing any additional constraints. Moreover, the new degree of freedom $\gamma$, associated with the connection, does not enter the cosmological field equations and therefore behaves as a spurious mode with no physical impact on the solutions. This feature has two important consequences. On the one hand, the number of effective degrees of freedom is reduced with respect to the other connections. On the other hand, the cosmological solutions are less constrained, since they are not subject to the additional condition arising from the connection field equations.

This situation changes for the cases referred to as Connection II and Connection III. In these realizations, the connection contributes an additional dynamical degree of freedom, and the cosmological system is supplemented by an extra equation arising from the condition $\mathcal{C}_t=0$. It should be stressed that this equation is not simply a consistency relation associated with the conservation of the energy--momentum tensor. Rather, it constitutes an independent field equation and must therefore be satisfied on an equal footing with the rest of the dynamical system. 

For this reason, a significant part of this work has been devoted to understanding how the connection field equation can be consistently incorporated into the system of modified Friedmann equations. The main result is that this equation can be interpreted as a constraint relating the quantities $\dot f_Q$ and $\dot f_C$, with the simplest realization corresponding to the condition $\dot f_Q=\dot f_C$. In this case, the general solution takes the form
\bea
f(Q,C)=g(Q+C)+\frac{k}{2}(Q-C)\; ,
\label{solCII-conclusion}
\eea
which plays a particularly important role because it defines a class of theories dynamically equivalent to $f(\mathring{R})$ gravity. The origin of this equivalence can be understood directly from the structure of Eq.~\eqref{solCII-conclusion}.

This result is not entirely surprising. Indeed, functions of the form $g(Q+C)$ are expected to lead to the same field equations for the three connection realizations, since such models can be rewritten as $g(\mathring{R})$ theories, for which only a single set of field equations exists. The presence of the nonhomogeneous term $k(Q-C)/2$ may appear more peculiar at first sight. However, because it is linear in both $Q$ and $C$, its contribution to the action can be recast as the Symmetric Teleparallel Equivalent of General Relativity (STEGR). Consequently, this term also preserves the dynamical equivalence with $f(\mathring{R})$ gravity.

As a consequence, the three cosmological systems associated with the different connection branches become dynamically indistinguishable. Rather than representing three independent cosmological sectors, they define a single degenerate cosmological sector of $f(Q,C)$ gravity, in which geometrically distinct affine connections give rise to identical cosmological dynamics. This degeneracy is lifted only when the integration constant appearing in the connection field equation is nonzero, making the choice of connection physically relevant.

As illustrative examples, we have analyzed both polynomial and exponential models. In both cases, the theory admits vacuum de Sitter solutions, showing that accelerated expansion can arise naturally from the gravitational sector. However, important differences emerge in the vacuum structure. While Minkowski spacetime is recovered as a vacuum solution in the polynomial case, this is no longer true when exponential corrections are introduced. In the latter case, the existence of a Minkowski solution requires the inclusion of a cosmological constant, highlighting the nontrivial impact of the exponential sector on the low-curvature regime of the theory.

Furthermore, we have explored models beyond the degenerate cosmological sector of $f(Q,C)$ gravity, where the choice of connection becomes physically relevant and gives rise to distinct cosmological background equations. Importantly, this analysis has been carried out while consistently satisfying the connection field equation, thereby guaranteeing the conservation of the energy--momentum tensor.

As a representative example, we have obtained cosmological solutions for the model
\bea
f(Q,C)=Q+\alpha Q C=(1+\alpha C)Q\; ,
\eea
in which the boundary term induces a correction to the effective gravitational coupling. In this sense, the factor $(1+\alpha C)$ can be interpreted as a geometry-dependent modification of the gravitational interaction, providing a simple framework to investigate genuinely new phenomenology beyond the sector dynamically equivalent to $f(\mathring{R})$ gravity.

While the cosmological solutions associated with Connection I can be obtained analytically, the cases of Connection II and Connection III require a numerical treatment due to the increased complexity of the corresponding dynamical systems. This additional complexity originates from two distinct sources. First, the field equations explicitly depend on a free function, $\gamma$ (or $\gamma_*$), associated with the connection. Second, the connection field equation $\mathcal{C}_t=0$ is no longer satisfied identically and must therefore be solved simultaneously with the modified Friedmann equations.

As a consequence, the cosmological solutions depend not only on the usual cosmological parameters, but also on the initial conditions associated with the connection sector. For Connection II, the relevant parameter space is given by $(h_0,\;\Omega_m,\;\alpha,\;\alpha_0,\;\gamma_0,\;\gamma'_0)$, while for Connection III it becomes $(h_0,\;\Omega_m\;,\alpha,\;\alpha_0,\;\gamma_{0},\;\gamma'_{*0})$. Moreover, the resulting evolution is sensitive to the epoch at which the initial conditions are imposed, providing an additional source of phenomenological richness absent in the simpler Connection I scenario.
 
Finally, our numerical analysis indicates that the parameters $\alpha$ and $\alpha_0$ cannot take arbitrarily large values if the resulting cosmological evolution is to remain close to the $\Lambda$CDM and CPL scenarios. Nevertheless, suitable choices of these parameters, together with the remaining free parameters of the model, generate controlled positive or negative corrections to the Hubble function at different stages of the cosmological evolution. These results demonstrate that the connection-dependent sector of $f(Q,C)$ gravity possesses sufficient phenomenological flexibility to reproduce cosmological histories close to the standard model while allowing for small deviations that deserve further investigation through a dedicated observational analysis.

\acknowledgments
We thank Konstantinos Dialektopoulos, Javier Rubio and José Beltrán Jiménez for their enlightening discussions and valuable feedback. This article is based upon work from COST Action CA21136 Addressing observational tensions in cosmology with systematics and fundamental physics (CosmoVerse) supported by COST (European Cooperation in Science and Technology). I.~A. is supported by the Basque Government Grant No.~IT1628-22, by Grant No.~PID2021-123226NB-I00 (funded by MCIN/AEI/10.13039/501100011033, by ``ERDF A way of making Europe''), and by the Grant Juan de la Cierva funded by MICIU/AEI/10.13039/501100011033 and by “ESF+”.

\bibliography{bib}

\end{document}